\documentclass[prl,showpacs]{revtex4}
	
\newcommand{\beq}{\begin{eqnarray}}
\newcommand{\eeq}{\end{eqnarray}}

\usepackage{graphicx,amssymb}

\begin{document}
\title{Asymptotically scale-invariant occupancy of phase space makes the entropy $S_q$ extensive
}

\author{
Constantino Tsallis$^{1,2}$, Murray Gell-Mann$^1$ and Yuzuru Sato$^1$\footnote{tsallis@santafe.edu, mgm@santafe.edu, ysato@santafe.edu}
}
\address{
$^1$Santa Fe Institute,
1399 Hyde Park Road,
Santa Fe, New Mexico 87501,  USA \\
$^2$Centro Brasileiro de Pesquisas F\'\i sicas,\
Rua Xavier Sigaud 150, 
22290-180 Rio de Janeiro-RJ, Brazil
}
\date{\today}

\begin{abstract}
Phase space can be constructed for $N$ equal and distinguishable subsystems that could be (probabilistically) either {\it weakly} (or {\it ``locally"}) correlated (e.g., independent, i.e., uncorrelated), or {\it strongly} (or {\it globally}) correlated.  If they are locally correlated, we expect the Boltzmann-Gibbs entropy $S_{BG} \equiv -k \sum_i p_i \ln p_i$ to be {\it extensive}, i.e., $S_{BG}(N)\propto N $ for $N \to\infty$. In particular, if they are independent,   $S_{BG}$ is {\it strictly additive}, i.e., $S_{BG}(N)=N S_{BG}(1),\, \forall N$. However, if the subsystems are globally correlated, we expect, for a vast class of systems, the entropy $S_q\equiv k\,[1- \sum_i p_i^q]/(q-1)$  (with $S_1=S_{BG}$) for some special value of $q\ne1$ to be the one which extensive (i.e., $S_q(N)\propto N $ for $N \to\infty$). Another concept which is relevant is strict or asymptotic {\it scale-freedom} (or {\it scale-invariance}), defined as the situation for which all marginal probabilities of the $N$-system coincide or asymptotically approach (for $N \to\infty$) the joint probabilities of the $(N-1)$-system. If each subsystem is a binary one, scale-freedom is guaranteed by what we hereafter refer to as the  {\it Leibnitz rule}, i.e., the sum of two successive joint probabilities of the $N$-system coincides or asymptotically approaches the corresponding joint probability of the $(N-1)$-system. The kinds of interplay of these various concepts are illustrated in several examples. One of them justifies the title of this paper. 
We conjecture that these mechanisms are deeply related to the very frequent emergence, in natural and artificial complex systems, of scale-free structures and to their connections with nonextensive statistical mechanics.
\end{abstract}
\maketitle

\section{I - Introduction}

The entropy $S_q$ \cite{Gell-MannTsallis04,history,history2}  is defined through
\begin{equation}
S_q \equiv k\frac{1-\sum_{i=1}^W p_i^q}{q-1}  \;\;\;(q \in {\cal R}; \; S_1=S_{BG}\equiv -k \sum_{i=1}^W p_i \ln p_i) \,,
\end{equation}
where $k$ is a positive constant ($k=1$ from now on) and $BG$ stands for {\it Boltzmann-Gibbs}.
This expression is the basis of {\it nonextensive statistical mechanics} \cite{Tsallis}, a current generalization of $BG$ statistical mechanics. For $q \ne 1$, $S_q$ is nonadditive -- hence nonextensive -- in the sense that for a system composed of (probabilistically) {\it independent} subsystems, the total entropy differs from the sum of the entropies of the subsystems.  However, the system may have special probability correlations between the subsystems such that extensivity is valid, not for $S_{BG}$, but for $S_q$ with a particular value of the index $q \ne 1$. 
In this paper, we address the case where the subsystems are all equal and distinguishable.  Their correlations may exhibit a kind of scale-invariance. 
We may regard some of the  situations of correlated probabilities as related to the remark (see \cite{nonlineardynamics} and references therein) that $S_q$ for $q \ne 1$ can be appropriate for nonlinear dynamical systems that have phase space unevenly occupied. We return to this point later.
   
We shall consider two types of models. The first one involves $N$ binary variables ($N=1,2,3,...$), and the second one involves $N$ continuous variables ($N=1,2,3$). In both cases, certain correlations that are scale-invariant in a suitable limit can create an intrinsically inhomogeneous occupation of phase space. Such systems are strongly reminiscent of the so called scale-free networks \cite{WattsBarabasi}, with their hierarchically structured hubs and spokes and their nearly forbidden regions. 

\section{II - Discrete models}

\subsection{Some basic concepts}

The most general probabilistic sets for $N$ equal and distinguishable binary subsystems are given in Table I with
\begin{equation}
\sum_{n=0}^N \frac{N!}{(N-n) !\,n!}\,\pi_{N,n}=1 \;\;\;\;(\pi_{N,n} \in [0,1]; \, N=1,2,3,... ;\, n=0,1,...,N)
\end{equation}

\begin{table}[htbp]
\begin{flushleft}
~~~~~~~~~~~~~~~~~~~~~~~~~~~~~~~~~~~~~$(N=0)$        ~~~~~~~~~~~~~~~~~~~~~~         ~~~~$1$~~~~\\
~~~~~~~~~~~~~~~~~~~~~~~~~~~~~~~~~~~~~$(N=1)$        ~~~~~~~~~~~~~~~~~~~         ~~~$\pi_{10}$~~$\pi_{11}$~~~\\ 
~~~~~~~~~~~~~~~~~~~~~~~~~~~~~~~~~~~~~$(N=2)$        ~~~~~~~~~~~~~~~~~         ~~$\pi_{20}$~~$\pi_{21}$~~$\pi_{22}$~~\\ 
~~~~~~~~~~~~~~~~~~~~~~~~~~~~~~~~~~~~~$(N=3)$        ~~~~~~~~~~~~~~~         ~$\pi_{30}$~~$\pi_{31}$~~$\pi_{32}$~~$\pi_{33}$~\\ 
~~~~~~~~~~~~~~~~~~~~~~~~~~~~~~~~~~~~~$(N=4)$        ~~~~~~~~~~~~~         $\pi_{40}$~~$\pi_{41}$~~$\pi_{42}$~~$\pi_{43}$~~$\pi_{44}$\\
\end{flushleft}
\vspace{-0.5cm}
\caption{Most general sets of joint probabilities for $N$ equal and distinguishable binary subsystems.} 
\end{table}
Let us from now on call {\it Leibnitz rule} the following recursive relation:
\begin{equation}
\pi_{N,n}+\pi_{N,n+1} =\pi_{N-1,n}\,\;\;\;(n=0,1,...,N-1;\,N=2,3, ...).                       .
\end{equation}
This relation guarantees what we refer to as {\it scale-invariance} (or {\it scale-freedom}) in this paper. Indeed, it guarantees that, for any value of $N$, the associated {\it joint probabilities} $\{\pi_{N,n}\}$ produce {\it marginal probabilities} which coincide with $\{\pi_{N-1,n}\}$ . Assuming $\pi_{10}+\pi_{11}=1$, and taking into acount that the $N$-th row has one more element than the $(N-1)$-th row,  a particular model is characterised by giving {\it one} element for each row. We shall adopt the convention of specifying
the set $\{\pi_{N,0} \in [0,1]$, $\forall N \}$ . Everything follows from it. There are many sets $\{\pi_{N,0} \}$ that satisfy Eq. (3). Let us illustrate with a few simple examples:\\

\noindent
(i) $\pi_{N,0}=\frac{(2\,\pi_{10})^N}{N+1}$ ($0\le \pi_{10}\le1/2;\,N=1,2,3, ...$). We have that all $2^N$ states have nonzero probability if $0<\pi_{10}\le1/2$ . The particular case $\pi_{10}=1/2$ recovers the original Leibnitz triangle itself \cite{Polya}: see Table II. \\

\noindent
(ii) $\pi_{N,0}= (\pi_{10})^{N^\alpha}$ ($\alpha \ge 0$; $N=1,2,3 ...$). The $\alpha=1$ instance corresponds to independent systems, i.e.,  $\pi_{N,n}=(\pi_{10})^{N-n}(1- \pi_{10})^n \,.$   If $0<\pi_{10} <1$, then all $2^N$ states have nonzero probability. The $\alpha=0$ instance corresponds to $\pi_{N,0}= \pi_{10}$, $\pi_{N,n}=0\,\,(n=1,2,...,N-1)$ and $\pi_{N,N}=1-\pi_{10}$ . If $0<\pi_{10} <1$, then only two among the $2^N$ states have nonzero probability, $\forall N$, namely the states associated with $\pi_{N,0}$ and $\pi_{N,N}$. \\

\begin{table}[htbp]
$(N=0)$        ~~~~~~~~~~~~~~~~~~~~~~         ~~~~$(1,1)$~~~~~~~~~~~~~~~~~~~~~~~~~~~~~~~~~\\
$(N=1)$        ~~~~~~~~~~~~~~~~         ~~~$(1,1/2)$~~$(1,1/2)$~~~~~~~~~~~~~~~~~~~~~~~~~\\ 
$(N=2)$        ~~~~~~~~~~~~         ~~$(1,1/3)$~~$(2,1/6)$~~$(1,1/3)$~~~~~~~~~~~~~~~~~~\\ 
$(N=3)$        ~~~~~~~         ~$(1,1/4)$~~$(3,1/12)$~~$(3,1/12)$~~$(1,1/4)$~~~~~~~~~\\ 
$(N=4)$        ~~         $(1,1/5)$~~$(4,1/20)$~~$(6,1/30)$~~$(4,1/20)$~~$(1,1/5)$~~\\
\caption{The left numbers within the parentheses correspond to Pascal triangle. The right numbers correspond to the Leibnitz harmonic triangle ($d=N$).} 
\end{table}

We may relax the Leibnitz rule to some extent by considering those cases where the rule is satisfied only asymptotically, i.e.,
\begin{equation}
\lim_{N\to\infty}\frac{\pi_{N,n}+\pi_{N,n+1}}{\pi_{N-1,n}}=1 \,\;\;\;(n=0,1,2, ...).
\end{equation}
Such cases will be said to be not strictly but {\it asymptotically scale-invariant} (or {\it asymptotically scale-free}). This is, for a variety of reasons, the situation in which we are primarily interested. The main reason is that what vast classes of natural and artificial systems typically exhibit is not precisely power-laws, but behaviors {\it which only asymptotically become power-laws} (once we have corrected, of course, for any finite size effects). This is consistent with the fact that within nonextensive statistical mechanics $S_q$ is optimized by $q$-exponential functions (see \cite{Gell-MannTsallis04}, references therein, and \cite{duality}), which only asymptotically yield power-laws. It is consistent also with a new central limit theorem that has been recently conjectured \cite{Tsallis2004a,moyano} for specially correlated random variables. 

Let us now introduce a further concept, namely {\it $q$-describability}. A model constituted by $N$ equal and distinguishable subsystems will be called {\it $q$-describable} if a value of $q$ exists such as $S_q(N)$ is {\it extensive}, i.e., $\lim_{N\to\infty} \frac{S_q(N)}{N} < \infty$. If that special value of $q$ equals unity, this corresponds to the usual $BG$ universality class. If that value of $q$ differs from unity, we will have nontrivial universality classes. If the subsystems $\{A_i\}$ are not necessarily equal, the system is $q$-describable if an entropic index $q$ exists such that $\lim_{N\to\infty} \frac{S_q(A_1+A_2+...+A_N)}{\sum_{i=1}^NS_q(A_i)} < \infty$. It should be clear that we could equally well demand the extensivity of say $S_{2-q}$ (or even of $S_{Q(q)}$, where $Q(q)$ is some monotonically decreasing function of $q$ satisfying $Q(1)=1$) instead of that of $S_q$ . This would of course have the effect of having nontrivial solutions for $q>1$ whenever we had solutions for $q<1$ if the extensivity that was imposed was that of $S_q$.

Finally, let us point out that we might consider the subsystems of a probabilistic system to be either {\it strongly} (or {\it globally}) {\it correlated} or {\it weakly} (or {\it ``locally"}) {\it correlated}.    The trivial case of {\it independence}, i.e., when the subsystems are {\it uncorrelated}, is of course a particular case of weakly correlated.  Let us make these notions more precise. A system is weakly correlated if for every generic (different from zero and from unity) joint probability $\pi^{A_1+A_2+...+A_N}_{i_1,i_2,...,i_N}$ a set of individual probablities $\{ \pi^{A_r}_{i_r}\}$ exists such that $ \lim_{N\to\infty} \frac{\pi^{A_1+A_2+...+A_N}_{i_1,i_2,...,i_N}}{\prod_{r=1}^N \pi^{A_r}_{i_r}}=1$. Otherwise, the system is said strongly correlated. The particular case of independence corresponds to 
$\pi_{i_r}^{A_r}= \sum_{i_1,i_2,...,i_{r-1},i_{r+1},...,i_N}\pi_{i_1,i_2,...,i_N}^{A_1+A_2+...+A_N} \;(r=1,2,...,N)$. 
If the subsystems are equal and binary, this definition becomes as follows: a system is weakly correlated if, for generic $\pi_{N,n}$, a probability $p_0$ exists such that $ \lim_{N\to\infty} \frac{\pi_{N,n}}{p_0^{N-n}(1-p_0)^n}=1$. Otherwise the system is said to be strongly correlated. The particular case of independence corresponds to $p_0=\pi_{10}$. In the present sense, weakly correlated systems could also be thought  and referred to as {\it asymptotically uncorrelated}.
The interplay of scale-invariance, $q$-describability and global correlation is schematized in Fig. 1.

We have verified that all systems illustrated in (i) and (ii) above belong to the $q=1$ class (see examples in Fig. 2). We next address $q \ne 1$ systems.
\begin{figure}
\begin{center}
\includegraphics[scale=0.6]{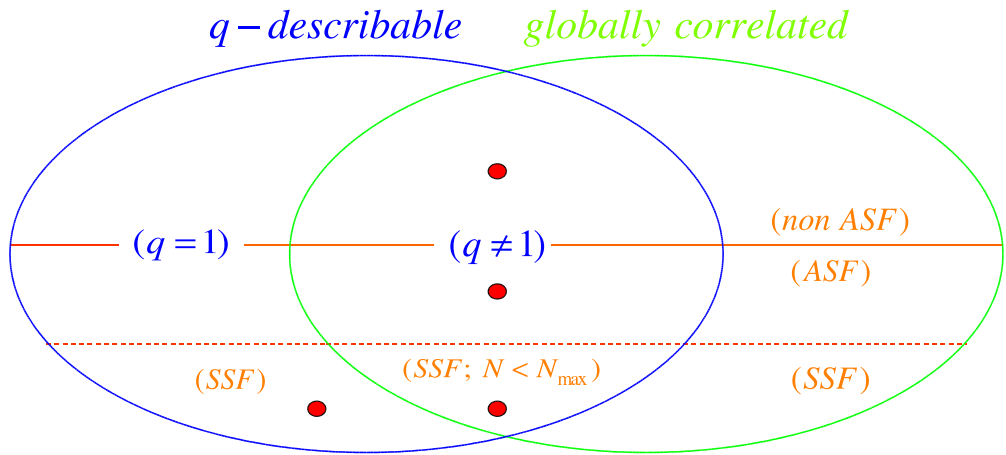}
\end{center}
\caption{\small Scheme representing the systems that are $q$-describable, globally correlated, asymptotically scale-free ($ASF$) and strictly scale-free ($SSF$). The $q=1$ region corresponds to ``locally" correlated systems. Leibnitz rule is strictly satisfied for $SSF$, but only asymptotically satisfied for $ASF$. Below  (above) the continuous red line we have the {\it ASF} ({\it non ASF}) systems. The $SSF$ systems (below the dashed red line) constitute a subset of the $ASF$ subset. The red spots correspond to the four families of discrete systems illustrated in the present paper: (a) $q \ne 1$ {\it non ASF} (upper spot; Eqs. (12) and (14)); (b) $q \ne 1$ {\it ASF} but {\it non SSF} (middle spot; Eqs. (17) and (24)); (c) $q \ne 1$ {\it SSF} (right bottom spot; Eq. (8)); (d) $q = 1$ {\it SSF} (left bottom spot; examples (i) and (ii) in the text). 
}
\end{figure} 

\begin{figure}
\begin{center}
\includegraphics[scale=0.56]{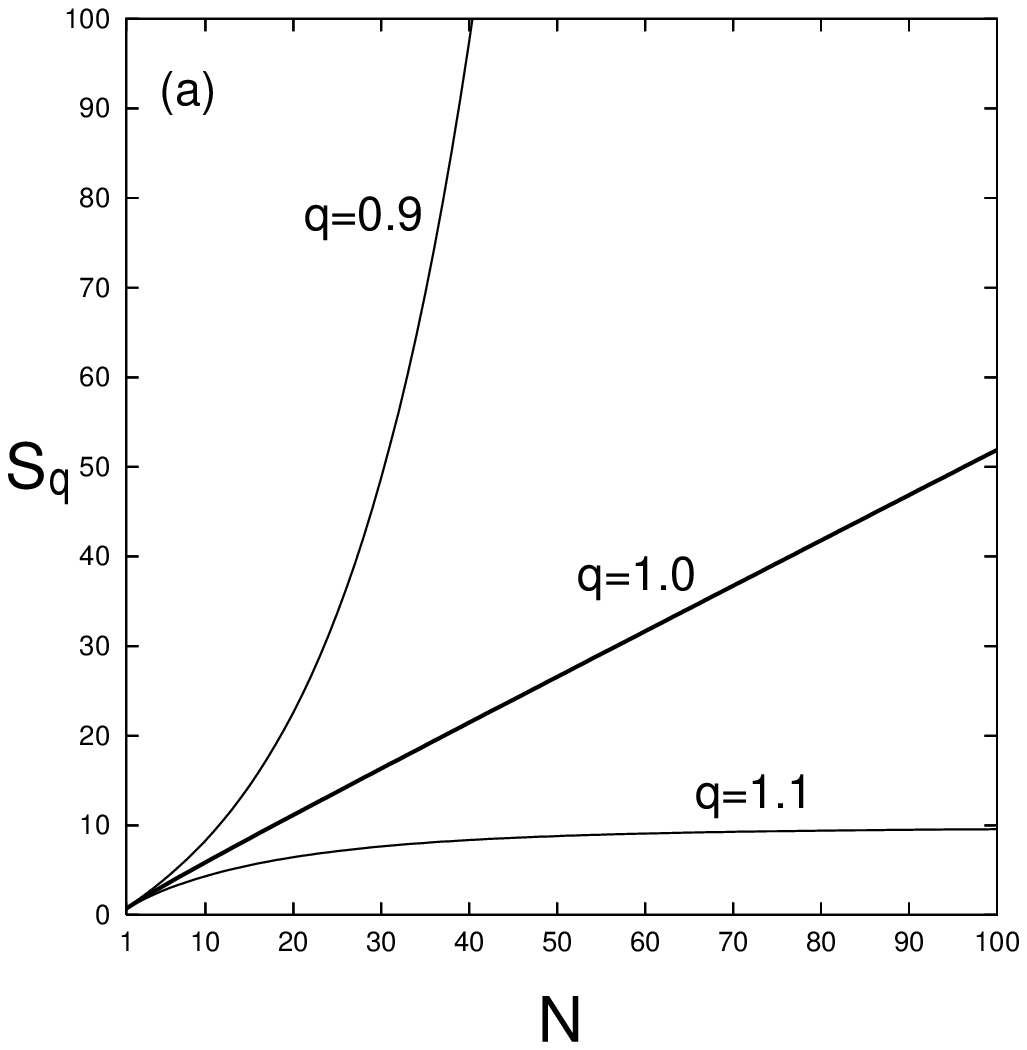}
\includegraphics[scale=0.56]{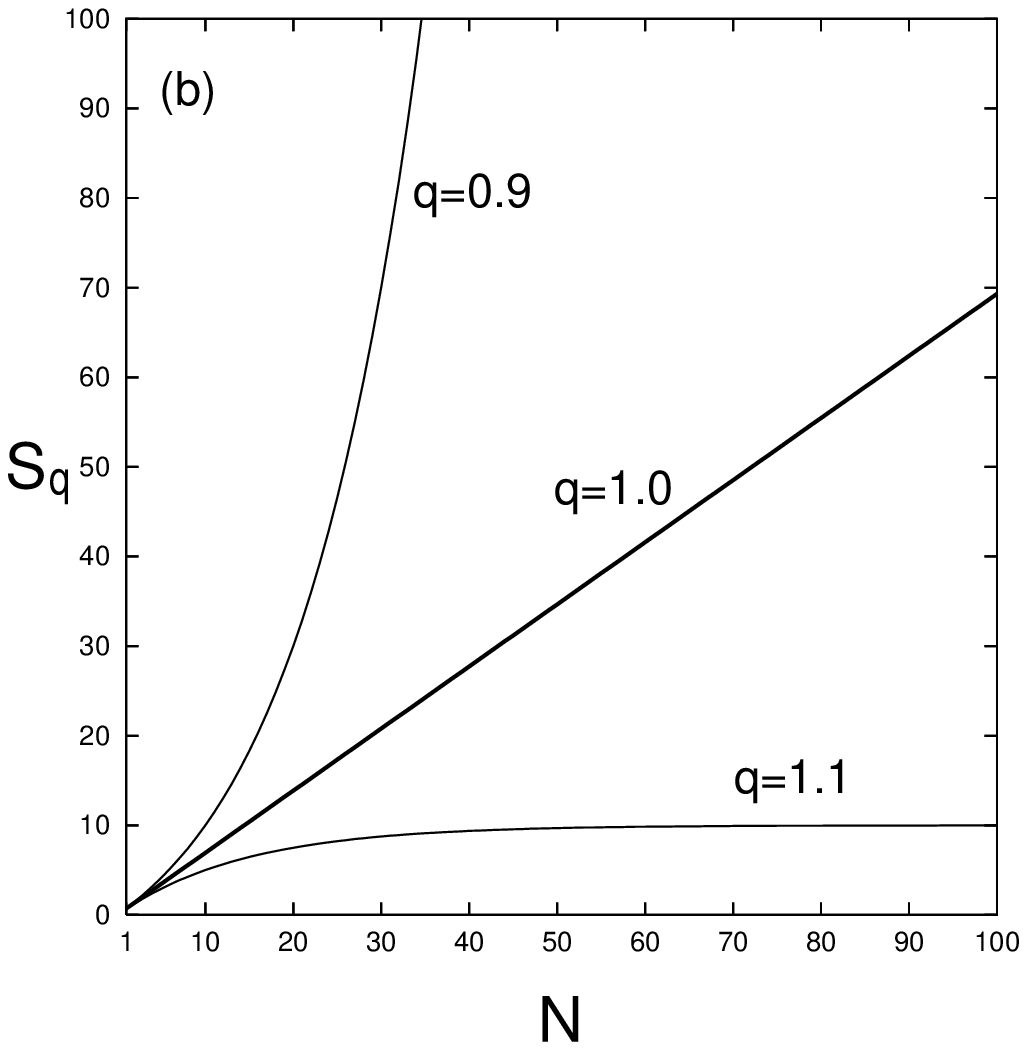}
\includegraphics[scale=0.56]{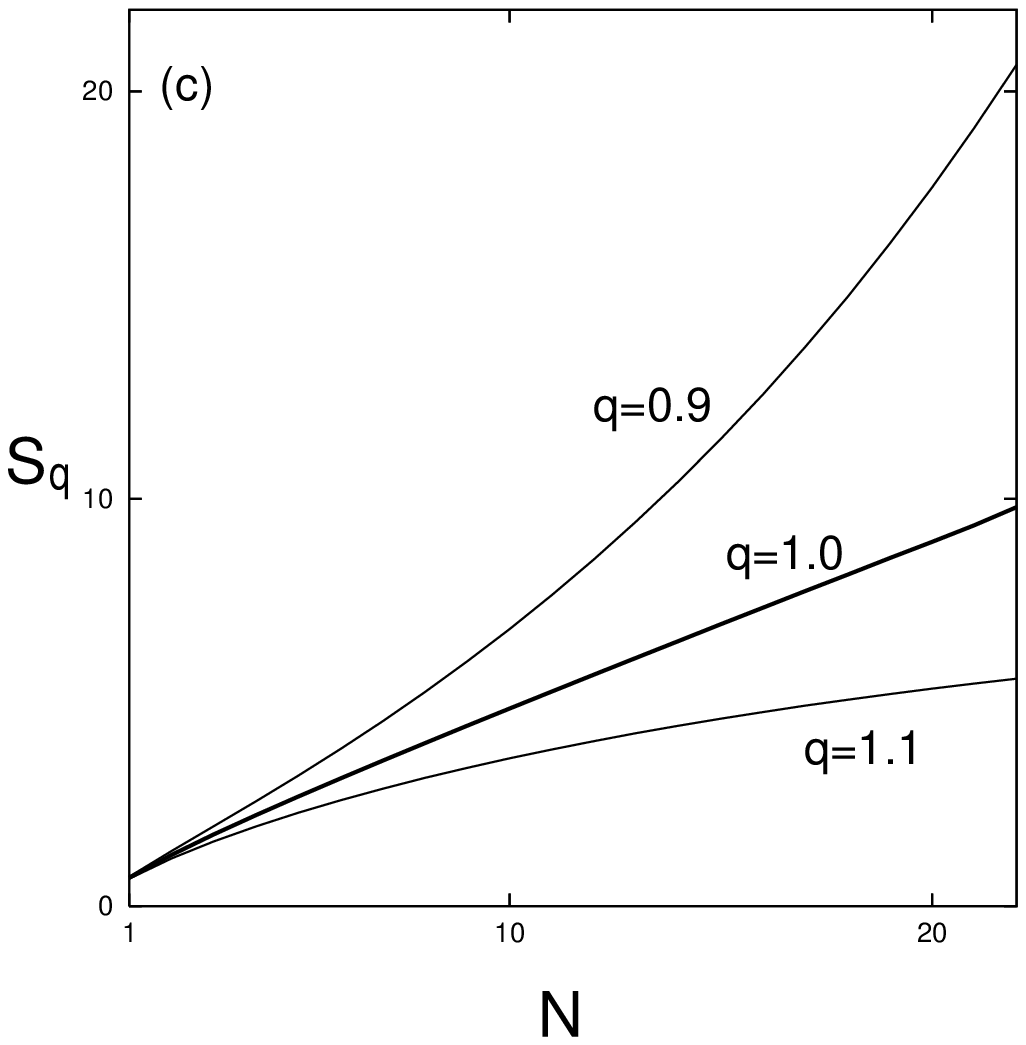}
\end{center}
\caption{\small $S_q(N)$ for (a) the Leibnitz triangle [the explicit expression $\pi_{N,n}=\frac{1}{(N+1)} \frac{(N-n)! \,n!}{N!}$ has been used to calculate $S_q(N)$], (b) $\alpha=1$ (i.e., independent subsystems) with $\pi_{10}=1/2$ [the explicit expression $\pi_{N,n}=(\pi_{10})^{N-n}(1- \pi_{10})^n$ has been used to calculate $S_q(N)$], and (c) $\alpha=1/2$ with $\pi_{10}=1/2$ [the recursive relation (3) has been used to calculated $S_q(N)$]. Only for $q=1$ we have a {\it finite} value for $\lim_{N\to\infty}S_q(N)/N$; it {\it vanishes} ({\it diverges}) for $q>1$ ($q<1$).
}
\end{figure}

\subsection{A strictly scale-invariant discrete model}

\begin{figure}
\begin{center}
\includegraphics[scale=0.8]{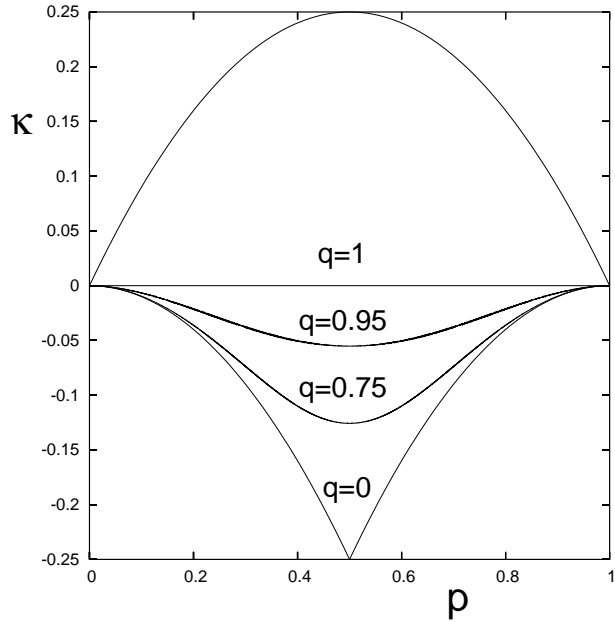}
\vspace{-0.5cm}
\end{center}
\caption{\small Curves $\kappa(p)$ which, for typical values of $q$, imply additivity of $S_q$. For $ -1/4 \le \kappa \le 0$ we have $\sqrt{-\kappa} \le p \le 1- \sqrt{-\kappa}$. For $0 \le \kappa \le 1/4$ we have $(1-\sqrt{1-4\kappa})/2 \le p \le (1+\sqrt{1-4\kappa})/2$ .
}
\label{fig2}
\end{figure}

In dealing with our first $q \ne 1$ {\it discrete} example, we start with two equal and distinguishable binary subsystems $A$ and $B$ ($N=2$). The associated joint probabilities are, with all generality, indicated in Table III, where $\kappa$ is the {\it correlation} \cite{covariance} between $A$ and $B$.
\begin{table}[htbp]
\begin{tabular}{c||c|c||cccccc}
$_A\setminus^B$    &  1                                    & 2                                                                                                                \\[1mm] \hline\hline
1  &  $p_{11}^{A+B} = p^2 +\kappa\;\;$                                  & $\;\;p_{12}^{A+B}=p\,(1-p)-\kappa\;\;$              & $\;\;p\;\;$   \\[3mm] \hline
2  &  $p_{21}^{A+B}=p\,(1-p)-\kappa    $                                  & $p_{22}^{A+B}=(1-p)^2+\kappa$                      & $1-p$       \\[3mm] \hline \hline
    &  $p$                                                            & $1-p$                                        & 1
\end{tabular}
\begin{tabular}{cccccccccccc}
& &&&&     \\[1mm] 
&&&&&&    \\[3mm] 
&&&&&&    \\[3mm] 
&&&&&& 
\end{tabular}
\begin{tabular}{c||c|c||c}
$_A\setminus^B$    &  1                                    & 2                                                                   \\[1mm] \hline\hline
1  &  $2p-1$                                  & $1-p$              & $\;\;p\;\;$   \\[3mm] \hline
2  &  $1-p$                                  & $0$                      & $1-p$       \\[3mm] \hline \hline
    &  $p$                                                            & $1-p$                                        & 1
\end{tabular}
\caption{{\it Left:} {\it Joint} and {\it marginal} probabilities for two binary subsystems $A$ and $B$.  Correlation $\kappa$ and probability $p$ are such that $0 \le p^2 +\kappa$, $p\,(1-p)-\kappa$, $(1-p)^2+\kappa \le 1$ ($\kappa=0$ corresponds to independence, for which case entropy additivity implies  $q=1$). {\it Right:} One of the two (equivalent) solutions for the particular case for which entropy additivity implies $q=0$. 
}
\end{table}
Let us now impose \cite{Tsallis2004b} additivity of $S_q$ \cite{observation}.
In other words, we choose $\kappa(p)$ such that $S_q(2)=2S_q(1) $, 
where (for $W=2$) $S_q(1)=\frac{1-p^q-(1-p)^q}{q-1}$,
and (for $W=4$) $S_q(2)=\frac{1- (p^2 +\kappa)^q -2[p\,(1-p)-\kappa]^q-[(1-p)^2+\kappa]^q}{q-1} $.
We focus on the solutions $\kappa_q(p)$ for $0 \le q \le1$  indicated in Fig. \ref{fig2} \cite{marsh}. 

With the convenient notation
\begin{eqnarray}
\pi_{10} \equiv r_{10} &\equiv& p_{1}^A                         = p    \nonumber  \\
\pi_{11} \equiv r_{01} &\equiv& p_{2}^{A}                         = (1-p) \nonumber \\
\pi_{20} \equiv   r_{20} &\equiv& p_{11}^{A+B}                         = p^2 +\kappa  \nonumber \\
\pi_{21} \equiv r_{11} &\equiv& p_{12}^{A+B} =p_{21}^{A+B} = p(1-p) -\kappa   \nonumber \\
\pi_{22} \equiv r_{02} &\equiv& p_{22}^{A+B}                         = (1-p)^2 +\kappa \,,
\end{eqnarray}
we can verify 
\begin{eqnarray}
r_{20}+2r_{11}+r_{02}&=&1\,,    \nonumber \\
r_{20} + r_{11}&=&r_{10} =p   \,, \nonumber \\
r_{11} + r_{02} &=& r_{01}=1-p \,.
\end{eqnarray}

Let us now address the case of three equal and distinguishable binary subsystems $A$, $B$ and $C$ ($N=3$). We present in Table IV probabilities that are {\it not} the most general ones, but rather general ones for which we have {\it strict scale invariance}, in the sense that {\it all} the associated marginal probability sets exactly reproduce the above $N=2$ case. Notice how strongly this construction reminds us of the one that occurs in the renormalization group procedures widely used in quantum field theory, the study of critical phenomena, and elsewhere \cite{Murray}.
\begin{table}[htbp]
\begin{tabular}{c||c|c||}
 $_A\setminus^B$    &  1                                                                & 2                                                    \\
[1mm] \hline\hline
1                              &  $p^3+\kappa_q(p) (2+p)$                            & $p^2(1-p)-\kappa_q(p) (1+p)$    \\   
                                & $[p^2(1-p)-\kappa_q(p) (1+p)]$                               &$[p(1-p)^2 +\kappa_q(p) \, p]$            \\                                               
[3mm] \hline
2                              &  $p^2(1-p)-\kappa_q(p) (1+p)$                                & $p(1-p)^2 +\kappa_q(p)\, p$             \\
                                &  $[p(1-p)^2 +\kappa_q(p)\, p]$                                      & $[(1-p)^3 + \kappa_q(p) (1-p)]$          \\                                         
[3mm] \hline \hline
\end{tabular}
\caption{Scale-invariant joint probabilities $p^{A+B+C}_{ijk}$ ($i,j,k=1,2$): the quantities without (within) square-brackets $[\;]$ correspond to state 1 (state 2) of subsystem $C$. 
}
\end{table}

With the convenient notation $\pi_{30} \equiv r_{30} \equiv p_{111}^{A+B+C}; \, \pi_{31} \equiv r_{21} \equiv p_{112}^{A+B+C}=p_{121}^{A+B+C}=p_{211}^{A+B+C}; \, \pi_{32} \equiv r_{12} \equiv     p_{221}^{A+B+C}=p_{212}^{A+B+C}=p_{122}^{A+B+C}; \, \pi_{33} \equiv r_{03} \equiv p_{222}^{A+B+C}$, and so on, 
we can verify
\begin{eqnarray}
r_{30}+3 r_{21} + 3 r_{12} + r_{03}&=&1 \,, \nonumber \\
r_{30}+r_{21}&=&r_{20}=p^2+\kappa_q(p) \,,  \nonumber \\
r_{21}+r_{12}&=&r_{11}=p(1-p) - \kappa_q(p) \,,  \nonumber \\
r_{12}+r_{03} &=&r_{02}=(1-p)^2 + \kappa_q(p) \,,
\end{eqnarray}
and so on.
\begin{table}[htbp]
\begin{flushleft}
~~~~~~~~~~~~~~~~~~~~~~~~~~~~~~~~~~~~~$(N=0)$        ~~~~~~~~~~~~~~~~~~~~~~         ~~~~$(1,1)$~~~~\\
~~~~~~~~~~~~~~~~~~~~~~~~~~~~~~~~~~~~~$(N=1)$        ~~~~~~~~~~~~~~~~         ~~~$(1,r_{10})$~~$(1,r_{01})$~~~\\ 
~~~~~~~~~~~~~~~~~~~~~~~~~~~~~~~~~~~~~$(N=2)$        ~~~~~~~~~~~~         ~~$(1,r_{20})$~~$(2,r_{11})$~~$(1,r_{02})$~~\\ 
~~~~~~~~~~~~~~~~~~~~~~~~~~~~~~~~~~~~~$(N=3)$        ~~~~~~~         ~$(1,r_{30})$~~$(3,r_{21})$~~$(3,r_{12})$~~$(1,r_{03})$~\\ 
~~~~~~~~~~~~~~~~~~~~~~~~~~~~~~~~~~~~~$(N=4)$        ~~~         $(1,r_{40})$~~$(4,r_{31})$~~$(6,r_{22})$~~$(4,r_{13})$~~$(1,r_{04})$\\
\end{flushleft}
\vspace{-0.5cm}
\caption{Merging of Pascal triangle with the present Leibnitz-like probability set. The particular case $r_{10}=r_{01}=1/2; \,
r_{20}=r_{02}=1/3 ;\; r_{11}=1/6; \, 
r_{30}=r_{03}=1/4; \; r_{31}=r_{13}=1/12 ; \,
r_{40}=r_{04}=1/5; \; r_{31}=r_{13}=1/20; \;r_{22}=1/30$, ..., recovers the Leibnitz triangle \cite{Polya}.
}
\end{table}

Let us complete this example by considering the generic case (arbitrary $N$). The results are presented in Table V, where we have merged
the Pascal triangle and the present Leibnitz-like triangle \cite{Polya}.  For the {\it left} elements, we have the usual Pascal rule, i.e., every element of the $N$-th line equals the sum of its ``north-west" plus its ``north-east"elements. For the {\it right} elements we have the property that every element of the $N$-th line equals the sum of its``south-west" plus its ``south-east" elements. 
In other words, for $(N=1,2,3,...; \,n=0,1,2,...,N)$, we have that $r_{N-n,n} + r_{N-n-1,n+1} = r_{N-n-1,n}$, and also that   $\sum_{n=0}^N \frac{N!}{(N-n)!\,n!} \,r_{N-n,n} =1\;\;\;\;(N=0,1,2,...)$. These two equations admit the following solution

\begin{eqnarray}
r_{N,0}& =&p^N + \kappa_q(p)     \frac{[N(1-p)+(p^N-1)]}{(1-p)^2}   \,, \nonumber    \\
r_{N-1,1} & =&p^{N-1}(1-p) - \kappa_q(p)\frac{1-p^{N-1}}{1-p} \,,     \\
r_{N-n,n} & =&p^{N-n}(1-p)^{n}\Bigr[1+\frac{\kappa_q(p)}{(1-p)^2} \Bigl] \;\;\;\; (2 \le n \le N) \,. \nonumber
\end{eqnarray}

Summarizing, as long as $r_{N,0} \ge 0$, this interesting structure takes automatically into account (i) the standard constraints of the theory of probabilities (nonnegativity and normalization of probabilities), and (ii) the scale-invariant structure which guarantees that {\it all the possible sets of marginal probabilities derived from the joint probabilities of $N$ subsystems reproduce the corresponding sets of joint probabilities of $N-1$ subsystems}. Consistently {\it $S_q$ is strictly additive} for all $N \le N_{max}$, where $N_{max}$ depends on $(p,q)$ \cite{marsh}. In this way, the correlation $\kappa_q(p)$ that we introduced between two subsystems will itself be  preserved for all $N \le N_{max}$.

Let us now address the following question: how {\it deformed}, and in what manner, is the occupation of the phase space ($N$-dimensional ``hypercube", in the same sense that the $N=2$ phase space may be seen as a ``square", and the $N=3$ one as a ``cube") in the presence of the scale-invariant correlation $\kappa_q(p)$ determined once and for all? (See Fig. 3) The most natural comparison is with the case of independence (which corresponds to $\kappa=0$, hence to $q=1$). It is then convenient to define the {\it relative discrepancy} $
\eta_{N-n,n} \equiv \{r_{N-n,n}/[p^{N-n}(1-p)^n]\}-1$ (naturally, other definitions for {\it discrepancy} can be used as well, but the present one is particularly simple). 
Since $n=0,1,2,...,N$, we may expect in principle to have $N+1$ {\it different} discrepancies. {\it It is not so!} Quite remarkably there are only {\it three} different ones, namely $\eta_{N,0}$, $\eta_{N-1,1}$, and all the others, which therefore coincide with $\eta_{0,N}$. They are given by
\begin{eqnarray}
\eta_{N,0}&=&  \frac{\kappa_q(p)}{(1-p)^2}   \Bigl[1+\frac{N(1-p)-1}{ p^{N}} \Bigr] \le 0  \,,\nonumber \\
\eta_{N-1,1} &=& \frac{\kappa_q(p)}{(1-p)^2}   \Bigl(1- \frac{1}{p^{N-1}} \Bigr)    \ge 0      \,,   \\
\eta_{N-n,n} &=&  \frac{\kappa_q(p)}{(1-p)^2} \le 0 \;\; (2 \le n \le N) \nonumber \,,
\end{eqnarray}
where the inequalities hold  for $0 \le q < 1$, for which $\kappa_q(p) \le 0$. Of course, the equalities in (9) correspond to $q=1$ (i.e., $\kappa=0$). See Fig. 3. We see that, for arbitrary $N \ge2$, only three different types of vertices emerge in the $N-$dimensional hypercube. These can be characterized by the $(1,1,...,1)$ corner, the $N$ sites along each cartesian axis emerging from this corner, and all the others. As $N$ increases, the middle type predominates more and more, with increasingly uneven occupation of phase space. 

The present example corresponds to $\pi_{N,0}=r_{N,0}$ as given in Eq. (8). It is important to notice in this case that, for fixed $(p,q)$ such that $p<1$ and $q<1$, there is a maximal value of $N$, noted $N_{max}(p,q)$, for which the analytical expression for $r_{N0}$ in Eq. (8) is nonnegative. For $N>N_{max}$, we are obliged to consider $ r_{N,0}=0$, which, through application of the Leibnitz rule, leads to violations of the nonnegativity of {\it all} $r_{N-n,n}$. When this happens, of course the additivity of the entropy, i.e., $S_q(N)=NS_q(1)$, does not hold any more. Unless we have the trivial situation $q=1$ (for which entropic additivity holds for all $0 \le p \le 1$), the thermodynamic limit $N \to\infty$ imposes $p=1$ for $0 \le q < 1$. Indeed $N_{max}(1,q) \to\infty \,\forall q\in [0,1]$. For all other values of $p<1$ and $q<1$,  $N_{max}(p,q)$ is finite.

\begin{figure}
\begin{center}
\includegraphics[scale=0.50]{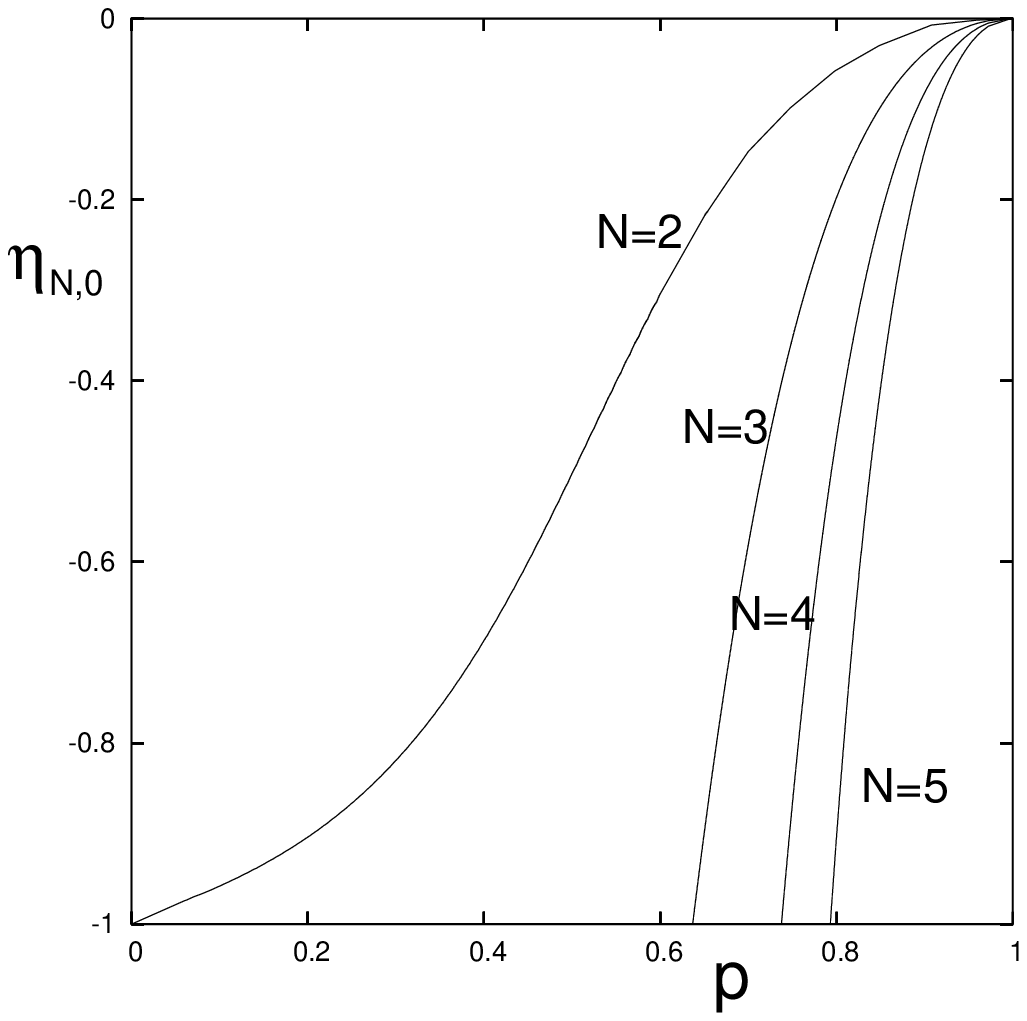}
\includegraphics[scale=0.50]{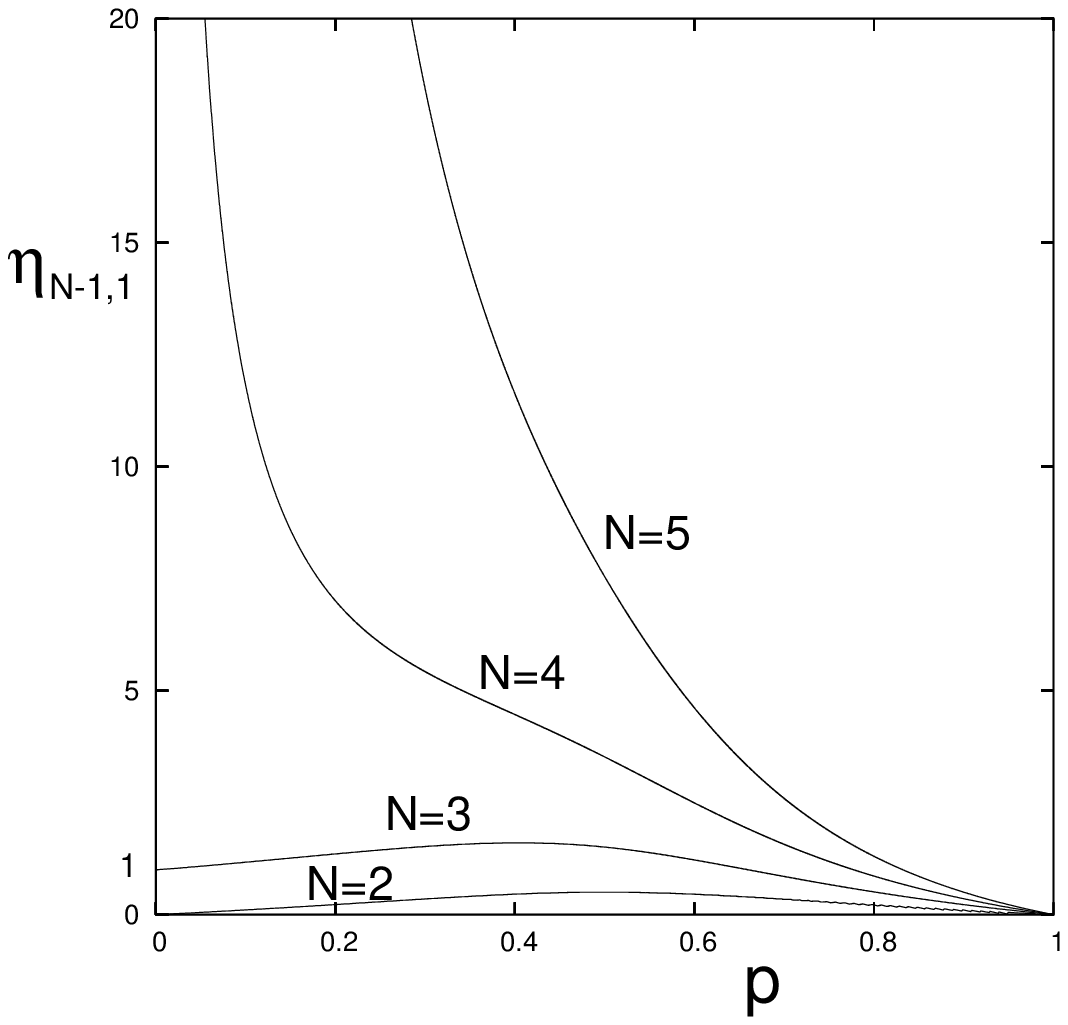}
\includegraphics[scale=0.50]{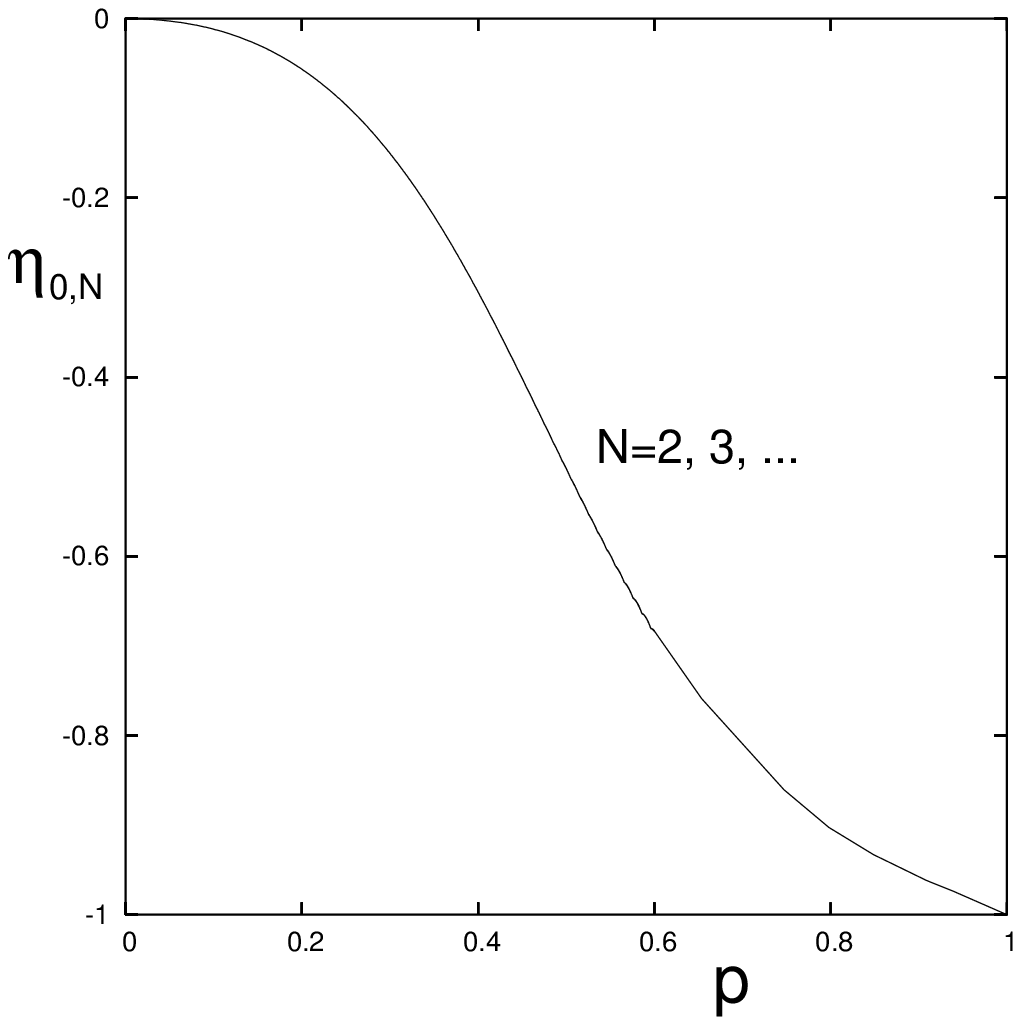}
\end{center}
\caption{\small $\eta_{N,0}(p)$ ({\it left}), $\eta_{N-1,1}(p)$ ({\it center}), and $\eta_{N-n,n}(p)$ ({\it right}), for $q=0.75$, and $N \le 5$. We see that, when $N$ increases, only the $N$ axes touching the $(1,1,...,1)$ corner of the hypercube remain occupied with an appreciable probability. Notice however that, for given $(p,q)$, $N$ is allowed to increase only up to a maximal value $N_{max}(p,q)$ (only $N_{max}(1,q)$ and $N_{max}(p,1)$ diverge).  
}
\end{figure}

\subsection{A discrete model that is not asymptotically scale-invariant}

Let us consider the probabilistic structure indicated in Table VI, where, for given $N$, only the $d+1$ first elements are different from zero, with $d=0,1,2,...,N$.

\begin{table}[h]
$(N=0)$ ~~~~~~~~~~~~~~~~~~~~~~~~~~~~~~$(1,1)$~~~~~~~~~~~~~~~~~~~~~~~~~~~~~~~~~~~~~~~~~~~~~~~~~~~$(1,1)$~~~~~~~~~~~~~~~~~~~~~~~~~~~~~~~~~~~~~~\\
$(N=1)$ ~~~~~~~~~~~~~~~~~~~~~~~$(1,\pi^{(1)}_{10})$~~$(1,\pi^{(1)}_{11})$~~~~~~~~~~~~~~~~~~~~~~~~~~~~~~~~~~$(1,\pi^{(2)}_{10})$~~$(1,\pi^{(2)}_{11})$~~~~~~~~~~~~~~~~~~~~~~~~~~~~~~~\\ 
$(N=2)$ ~~~~~~~~~~~~~~~~~$(1,\pi^{(1)}_{20})$~~$(2,\pi^{(1)}_{21})$~~$(1,0)$~~~~~~~~~~~~~~~~~~~~~~~~~$(1,\pi^{(2)}_{20})$~~$(2,\pi^{(2)}_{21})$~~$(1,\pi^{(2)}_{22})$~~~~~~~~~~~~~~~~~~~~~~~~~\\ 
$(N=3)$        ~~~~~~~~~~~~$(1,\pi^{(1)}_{30})$~~$(3,\pi^{(1)}_{31})$~~$(3,0)$~~$(1,0)$~~~~~~~~~~~~~~~~$(1,\pi^{(2)}_{30})$~~$(3,\pi^{(2)}_{31})$~~$(3,\pi^{(2)}_{32})$~~$(1,0)$~~~~~~~~~~~~~~~~~~~~~\\ 
$(N=4)$        ~~~~~~~$(1,\pi^{(1)}_{40})$~~$(4,\pi^{(1)}_{41})$~~$(6,0)$~~$(4,0)$~~$(1,0)$~~~~~~~$(1,\pi^{(2)}_{40})$~~$(4,\pi^{(2)}_{41})$~~$(6,\pi^{(2)}_{42})$~~$(4,0)$~~$(1,0)$~~~~~~~~~~~~~~~~~\\
\caption{Probabilistic models with $d=1$ ({\it left}) and $d=2$ ({\it right}).}
\end{table}
As we see, $\pi_{N,n}^{(d)}=0$ for $N \ge d+1$ and $n=d+1,d+2,...,N$. The total number of states is given by $W(N)=2^N$ ($\forall d)$, but the number of states with nonzero probability is given by 
\begin{equation}
W_{\mbox{\tiny {\it eff}}}(N,d)=\sum_{n=0}^d \frac{N!}{(N-n)!\,n!} \,,
\end{equation}
where {\it eff} stands for {\it effective}. For example, $W_{\mbox{\tiny {\it eff}}}(N,0)=1$, $W_{\mbox{\tiny {\it eff}}}(N,1)=N+1$, $W_{\mbox{\tiny {\it eff}}}(N,2)=\frac{1}{2}N(N+1)+1$, $W_{\mbox{\tiny {\it eff}}}(N,3)=\frac{1}{6}N(N^2+5)+1$, and so on. For fixed $d$ and $N\to\infty$ we have that
\begin{equation}
W_{\mbox{\tiny {\it eff}}}(N,d) \sim \frac{N^d}{d!}
\end{equation}

Let us now make a simple choice for the nonzero probabilities, namely {\it equal probalities}. In other words,
\begin{eqnarray}
\pi^{(d)}_{N,n}&=&1/2^N \;\;\;(\mbox{if}\;\;N \le d) \,,\nonumber \\
\pi^{(d)}_{N,n}&=& \frac{1}{W_{\mbox{\tiny {\it eff}}}(N,d)} \;\;\; (\mbox{if}\;\;N>d\;\;\mbox{and}\;\;n \le d) \,,\\
\pi^{(d)}_{N,n}&=& 0 \;\;\; (\mbox{if}\;\;N>d\;\;\mbox{and}\;\;n > d) \,.\nonumber 
\end{eqnarray}
See Table VII for an illustration of this model. 
\begin{table}[h]
$(N=0)$ ~~~~~~~~~~~~~~~~~~~~~~~~~~~~~~$(1,1)$~~~~~~~~~~~~~~~~~~~~~~~~~~~~~~~~~~~~~~~~~~~~~~~~~~~$(1,1)$~~~~~~~~~~~~~~~~~~~~~~~~~~~~~~~~~~~~~~\\
$(N=1)$ ~~~~~~~~~~~~~~~~~~~~~~~$(1,1/2)$~~$(1,1/2)$~~~~~~~~~~~~~~~~~~~~~~~~~~~~~~~~~~$(1,1/2)$~~$(1,1/2)$~~~~~~~~~~~~~~~~~~~~~~~~~~~~~~~\\ 
$(N=2)$ ~~~~~~~~~~~~~~~~~$(1,1/3)$~~$(2,1/3)$~~$(1,0)$~~~~~~~~~~~~~~~~~~~~~~~~~$(1,1/4)$~~$(2,1/4)$~~$(1,1/4)$~~~~~~~~~~~~~~~~~~~~~~~~~\\ 
$(N=3)$        ~~~~~~~~~~~~$(1,1/4)$~~$(3,1/4)$~~$(3,0)$~~$(1,0)$~~~~~~~~~~~~~~~~$(1,1/7)$~~$(3,1/7)$~~$(3,1/7)$~~$(1,0)$~~~~~~~~~~~~~~~~~~~~~\\ 
$(N=4)$        ~~~~~~~$(1,1/5)$~~$(4,1/5)$~~$(6,0)$~~$(4,0)$~~$(1,0)$~~~~~~~$(1,1/11)$~~$(4,1/11)$~~$(6,1/11)$~~$(4,0)$~~$(1,0)$~~~~~~~~~~~~~~~~~\\
\caption{Uniform distribution model with $d=1$ ({\it left}) and $d=2$ ({\it right}).}
\end{table}

The entropy for this model is given by
\begin{equation}
S_q(N)=\ln_q W_{\mbox{\tiny {\it eff}}}(N,d) \equiv \frac{ [W_{\mbox{\tiny {\it eff}}}(N,d)]^{1-q}-1}{1-q}   \sim \frac{N^{d(1-q)}}{(1-q)(d!)^{1-q}}\,,
\end{equation}
where we have used now Eq. (11). Consequently, $S_q$ is {\it extensive} (i.e., $S_q(N) \propto N$ for $N\to\infty$) if and only if
\begin{equation}
q=1-\frac{1}{d} \,.
\end{equation}
Hence, if $d=1,2,3...$, the entropic index monotonically approaches the $BG$ limit from below. 
We can immediately verify in Table VII (and using Eq. (12)) that this model violates the Leibnitz rule for all $N$, including asymptotically when $N\to \infty$. Consequently, it is neither strictly nor asymptotically scale-free. However, it is $q$-describable (see Fig. 1).

\subsection{An asymptotically scale-invariant discrete model}

Starting with the Leibnitz harmonic triangle,  
we shall construct a heterogeneous distribution $\pi^{(d)}_{N,n}$. The Leibnitz triangle is given in Table II and satisfies 
\begin{eqnarray}
p_{N,n}&=&p_{N+1,n}+p_{N+1,n+1},\\
p_{N,n}&=&\frac{1}{(N+1)} \frac{(N-n)! \,n!}{N!}.
\end{eqnarray}

We now define 
\begin{equation}
\pi^{(d)}_{N,n} \equiv \left\{
\begin{array}{ll}
p_{N,n}+l^{(d)}_{N,n} \, s^{(d)}_N&(n\le d)\\
0&(n>d)\\
\end{array}
\right.
\end{equation}
where the {\it excess probability} $s^{(d)}_N$ and the {\it distribution ration} $l^{(d)}_{N,n}$ (with $0<\epsilon<1$) are defined through
\begin{eqnarray}
s^{(d)}_N  &\equiv& \sum_{k=d+1}^N p_{N,k}=\frac{N-d}{N+1}\\
l^{(d)}_{N,n} &\equiv& \left\{
\begin{array}{ll}
1-\epsilon&(n=0)\\
(1-\epsilon) \,\epsilon^n \, \gamma_{N,n}^{(d)} \frac{(N-n)! \, n!}{N!}
&(0 < n < d)\\
\epsilon^d \, \gamma_{N,d}^{(d)}
&(n=d)\\
\end{array}
\right.
\end{eqnarray}
with
\begin{equation}
\gamma_{N,n}^{(d)} \equiv \prod_{k=1}^n\frac{1}{W_{\mbox{\tiny
eff}}(N,d)-W_{\mbox{\tiny eff}}(N,n-1)}
=\prod_{k=1}^n\frac{1}{\sum_{k=n}^d~  [N!/(N-n)! \,n!]}~~(n>0) \,,
\end{equation}
where  $W_{\mbox{\tiny {\it eff}}}(N,d)$ is given by Eq. (10).

\begin{table}[htbp]
$(N=0)$~~~~~~~~~~~~~~~~~~~~~~~~~~$(1,1)$~~~~~~~~~~~~~~~~~~~~~~~~~~~~~~~~~~~~~~~~~~~~~~~~~~~~~$(1,1)$~~~~~~~~~~~~~~~~~~~~~~~~~~~~~~~~~~~~~~~~~\\
$(N=1)$~~~~~~~~~~~~~~~~~~~$(1,1/2)$~~$(1,1/2)$~~~~~~~~~~~~~~~~~~~~~~~~~~~~~~~~~~~~~~$(1,1/2)$~~$(1,1/2)$~~~~~~~~~~~~~~~~~~~~~~~~~~~~~~~~~~  \\ 
$(N=2)$~~~~~~~~~~~~~~$(1,1/2)$~~$(2,1/4)$~~$(1,0)$~~~~~~~~~~~~~~~~~~~~~~~~~~~~~$(1,1/3)$~~$(2,1/6)$~~$(1,1/3)$~~~~~~~~~~~~~~~~~~~~~~~~~~~    \\ 
$(N=3)$~~~~~~~~$(1,1/2)$~~$(3,1/6)$~~$(3,0)$~~$(1,0)$~~~~~~~~~~~                                                                             ~~~~~~~~$(1,3/8)$~~$(3,25/288)$~~$(3,25/288)$~~$(1,0)$~~~~~~~~~~~~~~~~\\ 
$(N=4)$~~~$(1,1/2)$~~$(4,1/8)$~~$(6,0)$~~$(4,0)$~~$(1,0)$~~~~~~~~~~                                                                                              $(1,2/5)$~~$(4,21/400)$~~$(6,43/1200)$~~$(4,0)$~~$(1,0)$ ~~~~~~~~~~\\
\caption{Leibnitz-triangle-based $\epsilon=0.5$ probability sets: $d=1$ ({\it left}), and $d=2$ ({\it right}).} 
\end{table}

We have verified for $d=1,2,3,4$ and $N\to\infty$ a result that we expect to be correct for all $d<N/2$, namely that $0<\pi_{N,n+1}<< \pi_{N,n} \sim \pi_{N-1,n}<<1$, hence
\begin{eqnarray}
 \lim_{N\to\infty} &&\frac{\pi_{N-1,n}^{(d)}}{\pi_{N,n}^{(d)}+\pi_{N,n+1}^{(d)}}= 1 ~~,\\
 \lim_{N\to\infty} &&\frac{\pi_{N-1,d}^{(d)}}{\pi_{N,d}^{(d)}+0}=1 \,. 
\end{eqnarray}
In other words, the Leibnitz rule is asymptotically satisfied for the {\it entire} probability set $\{\pi_{N,n}\}$, i.e., this system has asymptotic scale invariance. Its entropy is given by 
\begin{eqnarray}
S_q(N,d)&=&\frac{1-\sum_{k=0}^{d}~ [N!/(N-n)!\,n!] [\pi^{(d)}_{N,k}]^q}{q-1}\,,
\end{eqnarray}
and we verify that a value of $q$ exists such that $ \lim_{N \to\infty }\frac{S_q(N, d)}{N}$ is finite. Our numerical results suggest that, for $0<\epsilon<1$, (see Fig. 5)
\begin{equation}
q=\frac{d-1}{d+1}\, . 
\end{equation}

\begin{figure}
\begin{center}
\includegraphics[scale=0.55]{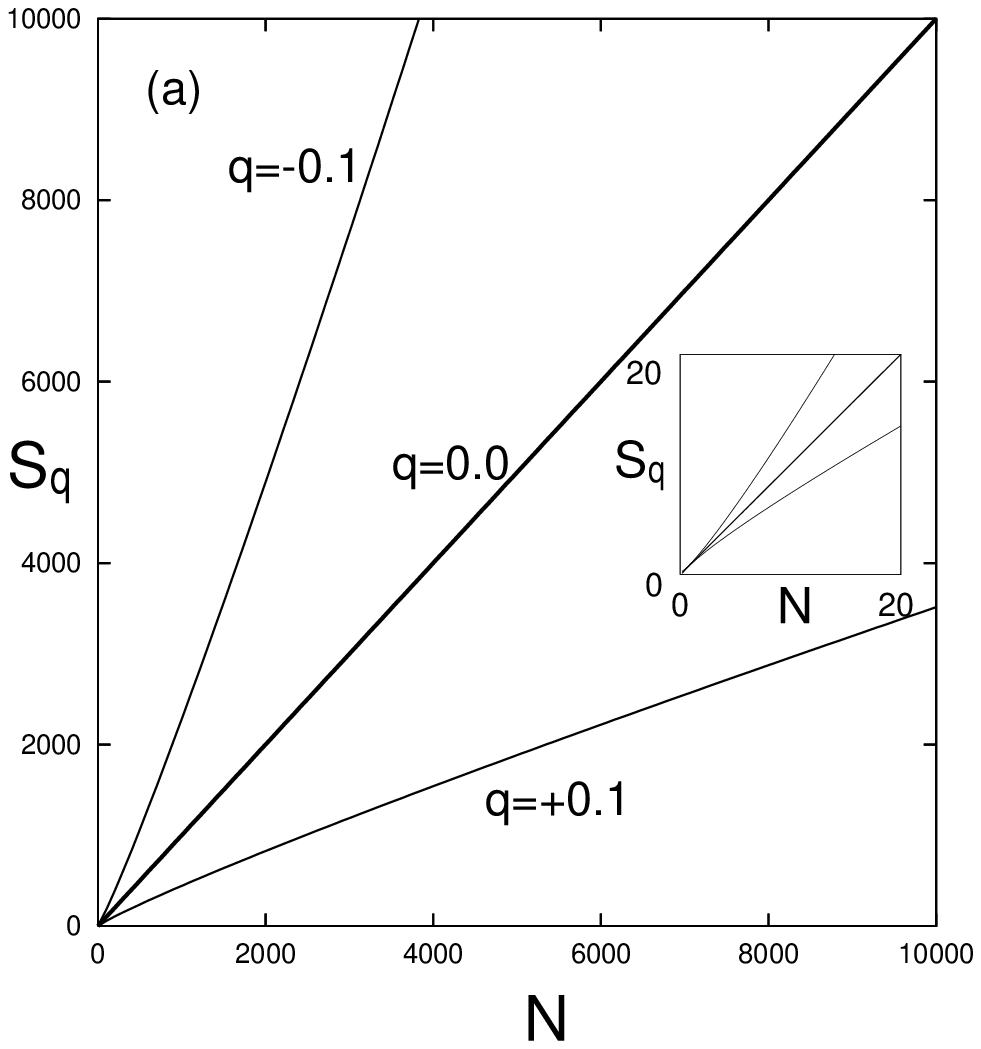}
\includegraphics[scale=0.563]{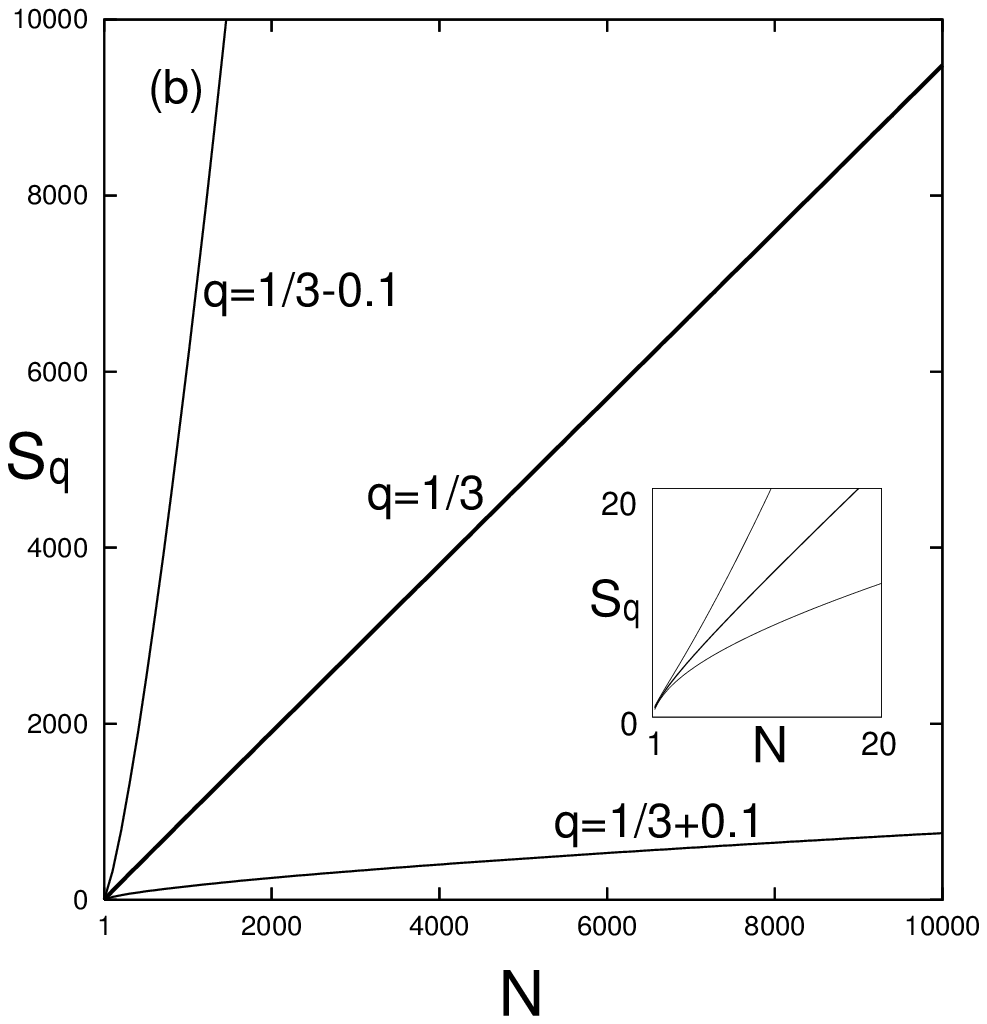}
\includegraphics[scale=0.552]{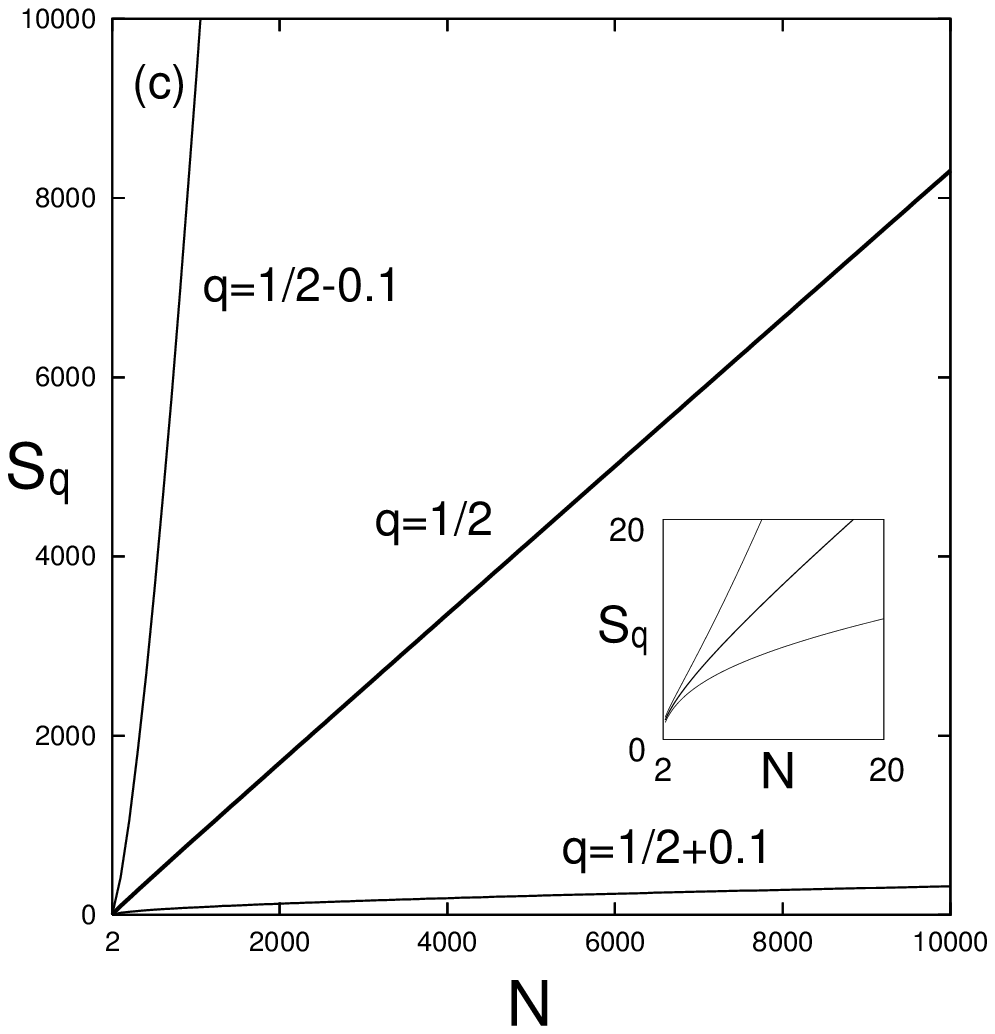}
\end{center}
\caption{\small Illustrations of the extensiviy of $S_q$ for the $q \ne 1$ $ASF$ model (with $\epsilon = 0.5$): (a) $d=1$; (b)  $d=2$; (c) $d=3$. Notice that the minimal value of $N$ equals $d-1$. {\it Insets:} Included to improve the perception of the fact that $\lim_{N\to\infty} \frac{S_q(N)}{N}$ {\it vanishes} ({\it diverges}) if $q>\frac{d-1}{d+1}$ ($q<\frac{d-1}{d+1}$), whereas it is {\it finite} for $q=\frac{d-1}{d+1}$.
}
\end{figure}

\section{III - Continuous model}

Let us now address our last example, namely a {\it continuous} model. It is known that classical mechanics violates the {\it 3rd principle of thermodynamics}, whereas quantum mechanics conforms to it. Indeed, in the latter we typically have $\lim_{T\to 0}\lim_{N \to \infty}S(N,T)/N=0$ ($T$ being the absolute temperature), whereas in the former such a limit is typically negative, and can even diverge to $-\infty$. Consistently, the present continuous model is going to have, as we shall see, difficulties of the same type. This, however, does not affect its scaling properties with $N$, which constitutes the central scope of the present paper. We shall therefore dedicate some effort to explore such continuous cases.
We consider the following probability distribution:
\begin{equation}
p(x) = \frac{2}{\sqrt{\pi}(2+a)} \, e^{-x^2}(1+ax^2)  \;\;\;\;( a \ge 0)
\end{equation}
We can verify that
$\int_{-\infty}^{\infty}dx \, p(x) = 1 \,.$ This distribution is illustrated in Fig.6.

\begin{figure}
\begin{center}
\includegraphics[scale=0.60]{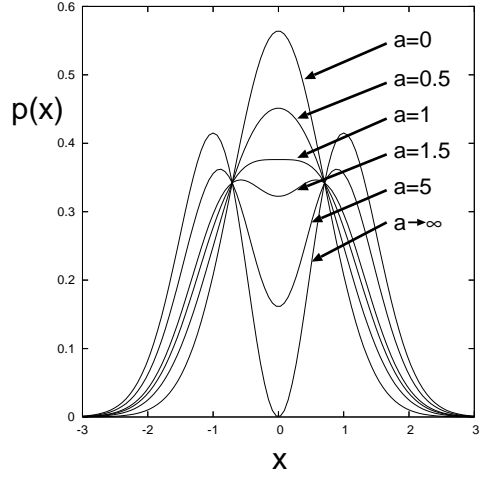}
\end{center}
\caption{\small Distribution $p(x)$ for typical values of $a$. The point shared by all distributions is located at $(|x|,p)=(1/\sqrt{2}\,,1/\sqrt{e\pi} ) \simeq (0.707,0.342)$.
}
\end{figure} 

The entropy corresponding to one subsystem (i.e., $N=1$) is given by

\begin{eqnarray}
S_q(1) &=& \frac{1-   \int_{-\infty}^{\infty} dx \, [p(x)]^q }{q-1}                \nonumber   \\
&=& \frac{1-  \Bigl[\frac{2}{\sqrt{\pi}(2+a)} \Bigr]^q  \int_{-\infty}^{\infty} dx \, e^{-q(x^2+y^2)}(1+ax^2)^q }{q-1}      \nonumber   \\
&=&\frac{1-    \frac{1}{\sqrt{q}}   \Bigl[\frac{2}{\sqrt{\pi}(2+a)} \Bigr]^q    I(a,q)   }{q-1}
\label{eqn12}
\end{eqnarray}
with \cite{Mathematica}
\begin{eqnarray}
I(a,q) &\equiv& \int_{-\infty}^{\infty} dz \, e^{-z^2}(1+\frac{a}{q}z^2)^q        \nonumber  \\
&=& \frac{\sqrt{\pi q} \, \Gamma(-\frac{1}{2}-q)  \,   _1F_1(\frac{1}{2},\frac{3}{2}+q,\frac{q}{a})}{\sqrt{a}  \, \Gamma(-q)} +\Bigl(\frac{a}{q} \Bigr)^q \,  \Gamma\Bigl(\frac{1}{2}+q\Bigr)\, _1F_1\Big(-q,\frac{1}{2}-q,\frac{q}{a}\Big) \,,
\end{eqnarray}
$\Gamma$ and $_1F_1$ being respectively the Riemann's $\Gamma$ and the hypergeometric functions. The $a$-dependence of $S_q$ for typical values of $q$ is depicted in Fig. (7). As expected for continuous distributions, negative values for $S_q$ do emerge.  

\begin{figure}
\begin{center}
\includegraphics[scale=1.2]{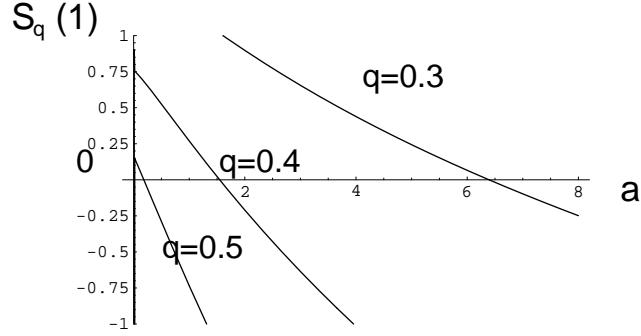}
\end{center}
\caption{\small Dependence of $S_q(1)$ on $a$ for typical values of $q$. $S_q$ is positive for $a<a_c(q)$ and negative for $a>a_c(q)$. The threshold value $a_c$ decreases from infinity to zero when $q$ increases from zero to unity. For $q=1$ we have that $S_{BG}<0$ for all $a>0$, thus exhibiting the well known difficulty of classical statistics. 
}
\end{figure} 

Let us now compose two such subsystems. If they are independent ($q=1$) we have
\begin{eqnarray}
P_1(x,y) =p(x)p(y) = \frac{4}{\pi(2+a)^2} \, e^{-(x^2+y^2)} \,      [1+a(x^2+y^2)+a^2x^2y^2]  
\end{eqnarray}
Of course,
$\int_{-\infty}^{\infty} \int_{-\infty}^{\infty} dx dy\, P_1(x,y) = 1$. 
For the general case, we propose the following simple generalization of $p(x)p(y)$:
\begin{eqnarray}
P_q(x,y) = \frac{4}{\pi(4+4A+B)}\, e^{-(x^2+y^2)} \,
    [1+A(x^2+y^2)+Bx^2y^2]  \,,
\end{eqnarray}
which satisfies
$\int_{-\infty}^{\infty} \int_{-\infty}^{\infty} dx dy\, P_q(x,y) = 1$. 
Of course, for $q=1$, we expect $(A,B)=(a,a^2)$.
Let us now calculate the marginal probability, i.e.,
\begin{eqnarray}
\int_{-\infty}^{\infty}  dy\, P_q(x,y) 
= \frac{2(2+A)   \,  e^{-x^2} }{\sqrt{\pi}(4+4A+B)}\Bigl[1+\frac{2A+B}{2+A}x^2\Bigr] 
\end{eqnarray}
We want this marginal probability to {\it recover} the original $p(x)$, so we impose
$(2A+B)/(2+A)=a$, 
which implies
$B=aA+2(a-A)$
and
$\int_{-\infty}^{\infty}  dy\, P_q(x,y) = p(x)$. It follows that
\begin{eqnarray}
P_q(x,y) = \frac{4}{\pi[4+2(a+A)+aA]} \,
e^{-(x^2+y^2)}  
 \{1+A(x^2+y^2)+[aA+2(a-A)]x^2y^2\}  \,.
\end{eqnarray}

Finally, to have $A$ as a function of $(q,a)$, we impose, as for the binary case,
\begin{equation}
S_q(2)=2S_q(1) \,,
\end{equation}
where $S_q(1)$ is given by Eq. (\ref{eqn12}) and
\begin{eqnarray}
S_q(2) &=&    \frac{1-   \int_{-\infty}^{\infty}  \int_{-\infty}^{\infty} dxdy \, [P_q(x,y)]^q }{q-1} \nonumber \\
&=&       \frac{1-  \Bigl[ \frac{4}{\pi[4+2(a+A)+aA]}  \Bigr]^q \int_{-\infty}^{\infty}  \int_{-\infty}^{\infty} dxdy \, e^{-q(x^2+y^2)}\{1+A(x^2+y^2)+[aA+2(a-A)]x^2y^2\}^q }{q-1}                 \nonumber         \\
&=& \frac{1-  \frac{1}{q} \Bigl[ \frac{4}{\pi[4+2(a+A)+aA]}  \Bigr]^q J(a,A,q) }{q-1}      
\end{eqnarray}
with \cite{Mathematica}
\begin{eqnarray}
J(a,A,q) &\equiv&      \int_{-\infty}^{\infty}  \int_{-\infty}^{\infty} du\,dv \, e^{-(u^2+v^2)}\Bigl[1+\frac{A}{q}(u^2+v^2)+   \frac{aA+2(a-A)}{q^2}u^2v^2\Bigr]^q \nonumber  \\
&=& \frac{1}{\Gamma(-q)} \int_{-\infty}^{\infty} dz \sqrt{\frac{1+(A/q)z^2}{(A/q)+[(aA+2(a-A))/q^2]z^2}}  \,   e^{-z^2}(1+(A/q)z^2)^q  \nonumber \\
 &\times&  \Bigl[ \sqrt{\pi} \, \Gamma\Bigl(\frac{1}{2}-q\Bigr) \, _1F_1\Bigl(\frac{1}{2},\frac{3}{2}+q,\frac{1+(A/q)z^2}{(A/q)+[(aA+2(a-A))/q^2]z^2} \Bigr) \nonumber \\
&&    \;\;\; + \; \Bigl(     \frac{(A/q)+[(aA+2(a-A))/q^2]z^2}{1+(A/q)z^2}  \Bigr)^{\frac{1}{2}+q} \Gamma(-q) \Gamma\Bigl(\frac{1}{2}+q \Bigr)   \nonumber \\
&&  \;\;\;\;\;\;\;\;\;\;\;\;\;\;\;\;\;  \times \;\;  _1F_1\Bigl(  -q, \frac{1}{2}-q, \frac{1+(A/q)z^2}{(A/q)+[(aA+2(a-A))/q^2]z^2}            \Bigr)                                              \Bigr]
\end{eqnarray}

See in Fig. 8 the $a$-dependence of $A$ for typical values of $q$. 
\begin{figure}
\begin{center}
\includegraphics[scale=0.6]{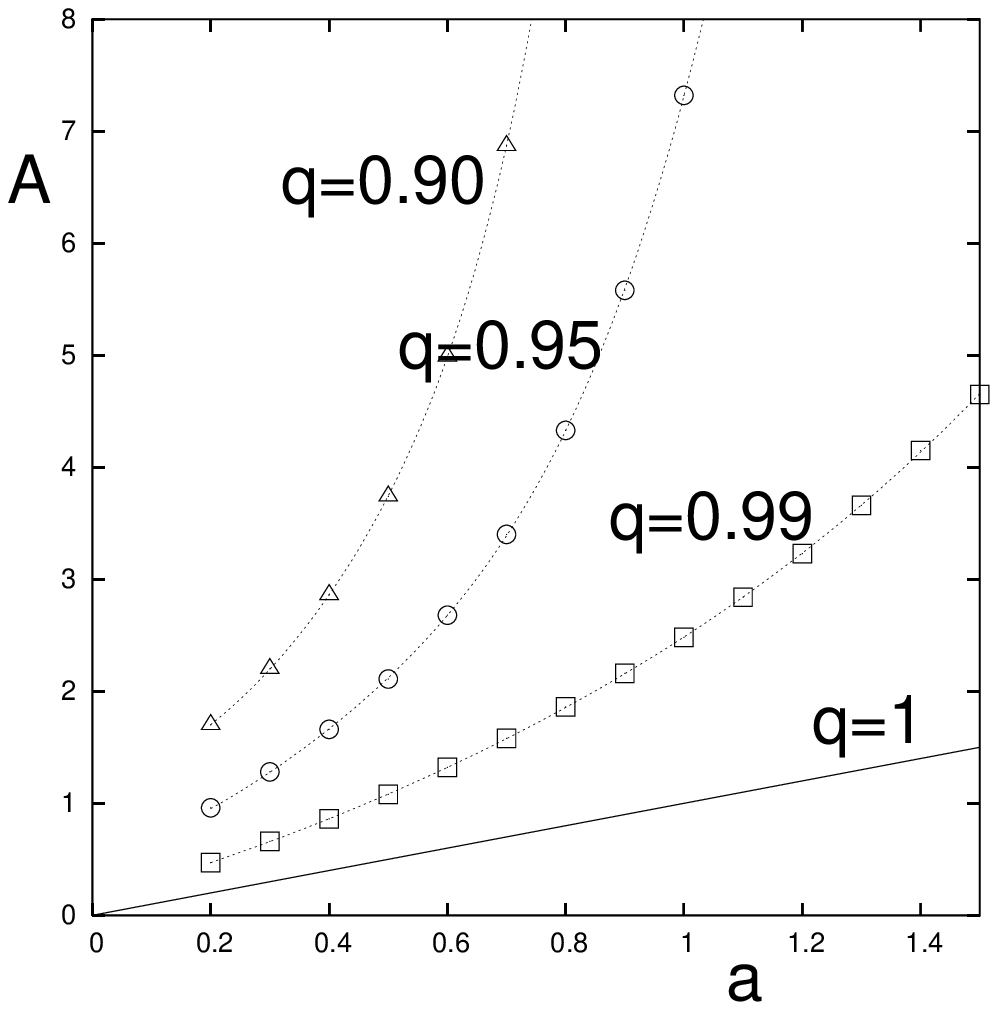}
\includegraphics[scale=0.6]{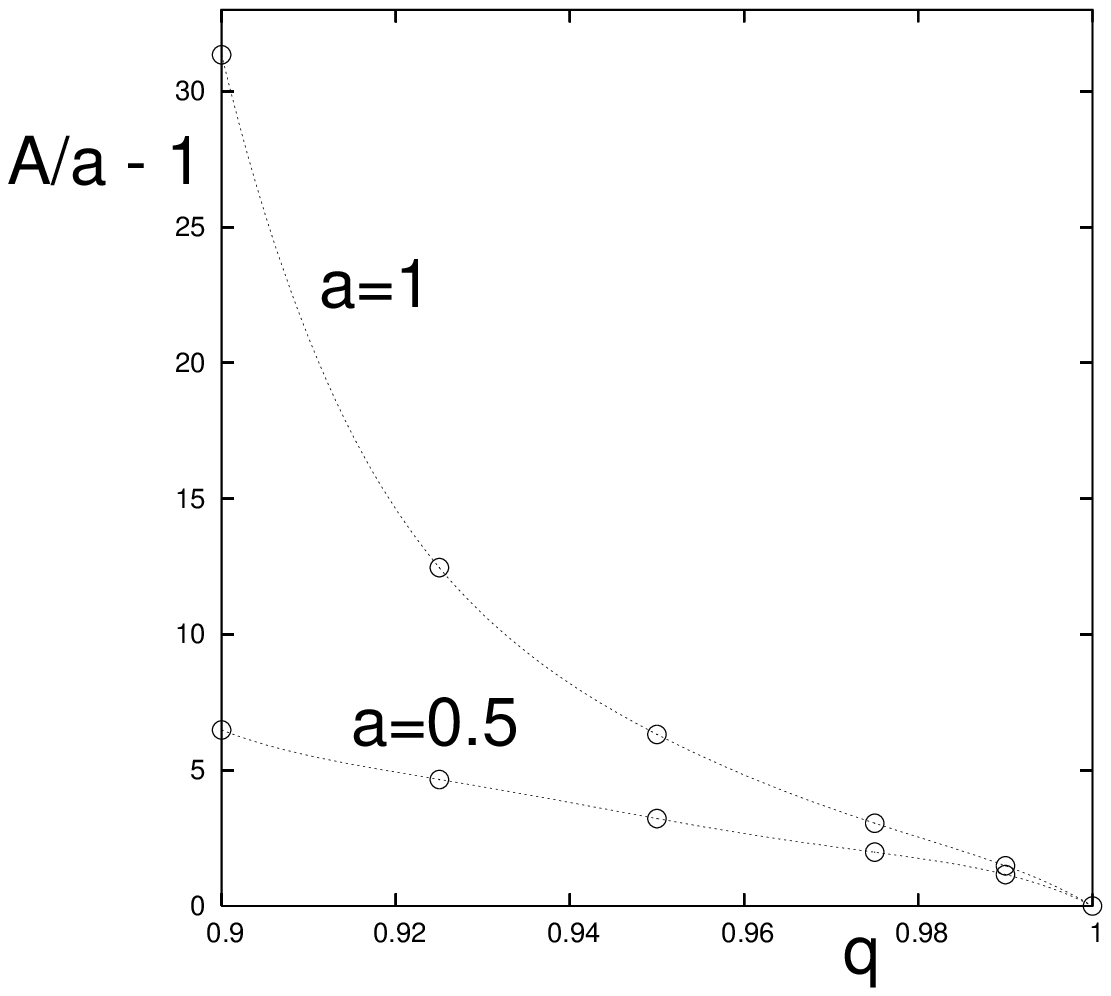}
\end{center}
\caption{\small $(a,q)$-dependence of $A$ ($A=a$ for $q=1$). {\it Left:} For typical values of $q$.  {\it Right:} For typical values of $a$.  
}
\end{figure} 
Finally, the {\it relative discrepancy} 
\begin{equation}
\eta(x,y) \equiv \frac{P_q(x,y)}{P_1(x,y)} -1
\end{equation}
is illustrated in Fig. 9 for a typical set  $(a,q)$.  
\begin{figure}
\begin{center}
\includegraphics[scale=1.0]{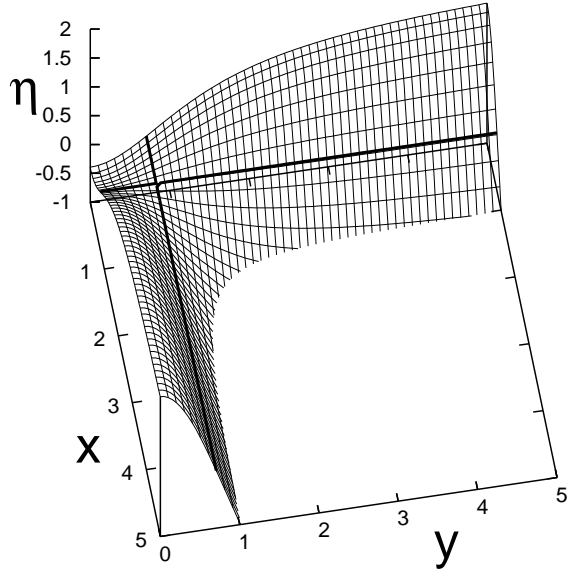}
\end{center}
\caption{\small $\eta(x,y;a,q)$ for  $(a,q)=(0.5,0.95)$ (hence $A=2.12$); $x=y$ is a plane of symmetry, i.e., $\eta(x,y;a,q)=\eta(y,x;a,q)$. The two bold straight lines correspond to $\eta=0$.
}
\end{figure} 
For higher values of $N$ we follow here a procedure similar to the one in our discrete example {\it SSF} of Fig. 1. Let us address the $N=3$ case. For the case of independence, we have 
\begin{equation}
P_1(x,y,z)=p(x)p(y)p(z) \propto e^{-(x^2+y^2+z^2)}[1+a(x^2+y^2+z^2)+a^2(x^2y^2+y^2z^2+z^2x^2)+a^3x^2y^2z^2].
\end{equation} 
We consistently assume
\begin{equation}
P_q(x,y,z) = \frac{8}{\pi^{3/2}(8+12A_3+6B_3+C_3)} e^{-(x^2+y^2+z^2)}[1+A_3(x^2+y^2+z^2)+B_3(x^2y^2+y^2z^2+z^2x^2)+C_3 x^2y^2z^2] \,, 
\end{equation}
which satisfies $\int_{-\infty}^{\infty} \int_{-\infty}^{\infty}\int_{-\infty}^{\infty}dx \,dy\, dz\, P_q(x,y,z)=1$. Clearly, for $q=1$,  $(A_3,B_3,C_3)=(a,a^2,a^3)$. For the general case, we impose that $\int_{-\infty}^{\infty}dzP_q(x,y,z)=P_q(x,y)$, i.e., the $N=2$ distribution as given by  Eq. (31). This imposition implies
\begin{eqnarray}
\frac{2A_3+B_3}{2+A_3}&=&A_2\equiv A \,, \nonumber \\
\frac{2B_3+C_3}{2+A_3}&=&B_2 \equiv B   \,,  \\
\frac{2+A_3}{ 8+12A_3+6B_3+C_3}&=& \frac{1}{4+4A_2+B_2} \,,                        \nonumber
\end{eqnarray}
hence
\begin{eqnarray}
A_3&=& \frac{4A_2-2B_2+C_3}{4-2A_2+B_2}  \,,                           \nonumber \\
B_3&=& \frac{4B_2+(A_2-2)C_3}{4-2A_2+B_2}   \,. 
\end{eqnarray}
The coefficient $C_3>0$ must satisfy that $C_3=a^3$ for $q=1$. If $S_q(3)=3 S_q(1)$ is automatically satisfied, we have some freedom for choosing $C_3$. Natural choices could be $C_3=a^3$ and $C_3=A_3B_3$ (which automatically satisfies $C_3=a^3$ for $q=1$). If, however, $S_q(3) \ne 3 S_q(1)$, we can impose the equality and determine a better approximation for $q$. The new value is expected to be only slightly different from the one that we already determined by imposing entropic additivity for $N=2$. The procedure can in principle be iteratively repeated for increasing $N$. Although such a study has its own interest, it lies outside the scope of this article.

\section{IV - Final remarks}

Let us now critically re-examine the physical entropy, a concept which is intended to measure the nature and amount of our ignorance of the state of the system. As we shall see, extensivity may act as a guiding principle. 
Let us start with the simple case of an isolated classical system with {\it strongly} chaotic nonlinear dynamics, i.e., at least one {\it positive} Lyapunov exponent. For almost all possible initial conditions, 
the system quickly visits the various admissible parts of a {\it coarse-grained} phase space in a virtually homogeneous manner. Then, when the system achieves {\it thermodynamic equilibrium}, our knowledge is as meager as possible ({\it microcanonical ensemble}), i.e., just the Lebesgue measure $W$ of the appropriate (hyper)volume in phase space (continuous degrees of freedom), or the number $W$ of possible states (discrete degrees of freedom). The entropy is given by $S_{BG}(N) \equiv k \ln W(N)$ ({\it Boltzmann principle} \cite{Einstein1910,Cohen2004}). 
If we consider independent equal subsystems, we have $W(N)=[W(1)]^N$, hence $S_{BG}(N)=NS_{BG}(1)$. If the $N$ subsystems are only {\it locally} correlated, we expect $W(N) \sim \mu^N \;(\mu \ge 1)$, hence $\lim_{N \to\infty}S_{BG}(N)/N=\mu$, i.e.,
the entropy is {\it extensive} (i.e., {\it asymptotically additive})

Consider now a strongly chaotic case for which we have more information, e.g., the set of probabilities $\{p_i\} \;(i=1,2,...,W)$ of the states of the system. The form $S_{BG}    \equiv  -k\sum_{i=1}^W p_i \ln p_i$ yields $S_{BG}(A+B)=S_{BG}(A)+S_{BG}(B)$ in the case of independence ($p_{ij}^{A+B}=p_i^Ap_j^B$). This form, although more general than $k \, ln W$ (corresponding to equal probabilities), still satisfies additivity. 
It frequently happens, though, that we do not know the {\it entire} set $\{p_i\}$, but only some constraints on this set, besides the trivial one $\sum_{i=1}^W p_i=1$. The typical case is Gibbs' canonical ensemble (Hamiltonian system in longstanding contact with a thermal bath), in which case we know  the mean value of the energy ({\it internal energy}). Extremization of $S_{BG}$ yields, as well known, the celebrated BG weight, i.e., $p_i \propto e^{-\beta E_i}$, with $\beta \equiv 1/kT$ and $\{E_i\}$ being the set of possible energies. This distribution recovers the microcanonical case (equal probabilities) for $T \to\infty$.

Let us now address more subtle physical systems (still within the class associated with strong chaos), namely those in which the particles are indistinguishable (bosons, fermions). This new constraint leads to a substantial modification of the description of the states of the system, and the entropy form has to be consistently modified, as shown in any textbook.  These expressions may be seen as further generalizations of $S_{BG}$, and the extremizing probabilities constitute, {\it at the level of the one-particle states}, generalizations of the just mentioned BG weight, recovered asymptotically at high temperatures. It is remarkable that, through these successive generalizations (and even more, since correlations due to local interactions might exist in addition to those connected with quantum statistics), {\it the entropy remains extensive}. 
Another subtle case is that of thermodynamic critical points, where correlations at all scales exist. There we can still refer to $S_{BG}$, but it exhibits singular behavior \cite{singular}. 

Finally, we address the completely different class of systems for which the condition of independence is severely violated (typically because the system is only {\it weakly chaotic}, i.e., its sensitivity to the initial conditions grows slowly with time, say as a {\it power-law}, with the maximal Lyapunov exponent vanishing). In such systems, long range correlations typically exist that unavoidably point toward generalizing the entropic functional, essentially because the effective number of visited states grows with $N$ as something like a power law instead of exponentially. 
We exhibited here  such examples  for which (either exact or asymptotic) {\it scale-invariant correlations} are present.
There the entropy $S_q$ for a special value of  $q \ne 1$ is {\it extensive},  whereas $S_{BG}$ is {\it not}. 

Weak departures from independence make $S_{BG}$ lose strict additivity, but not {\it extensivity}. 
Something quite analogous is expected to occur for scale-invariance in the case of $S_q$ for $q \ne 1$. Amusingly enough, we have shown (see also \cite{Tsallis2004a,Tsallis2004b}) that the ``nonextensive" entropy $S_q$ --- indeed nonextensive for independent subsystems --- {\it acquires extensivity in the presence of suitable asymptotically scale-invariant correlations}. Thus arguments presented in the literature that involve $S_q$ (with $q \ne 1$) concomitantly with the assumption of independence should be revisited. In contrast, those arguments based on extremizing $S_q$, without reference to the composition of probabilities, remain unaffected. 
While reference to ``nonextensive statistical mechanics" still makes sense, say for long-range interactions, we see that the usual generic labeling of the entropy $S_q$ for $q \ne 1$ as ``nonextensive entropy" can be misleading.   

The asymptotic scale invariance on which we focus appears to be connected with the asymptotically scale-free occupation of phase space that has been conjectured \cite{Gell-MannTsallis04} to be dynamically generated by the complex systems addressed by nonextensive statistical mechanics (see also \cite{Panchos}). {\it Extensivity} --- together with {\it concavity}, {\it Lesche-stability} \cite{Lesche}, and {\it finiteness of the entropy production per unit time} --- increases the suitability of the entropy $S_q$ for linking, with no major changes, statistical mechanics to thermodynamics.

Last but not least, the probability structure of our discrete cases is, interestingly enough, intimately related to both the Pascal and the Leibnitz triangles.
 
\section*{Acknowledgments}

We are grateful to R. Hersh for pointing out to us that the joint-probability structure of one of our discrete models is analogous to that of the Leibnitz triangle.  We have also benefited from very fruitful remarks by J. Marsh \cite{marsh} and L.G. Moyano \cite{moyano}. Support from SI International and AFRL is acknowledged as well. Finally, the work of one of us (M. G-M.) was supported by the C.O.U.Q. Foundation and by Insight Venture Management. The generous help provided by these organizations is gratefully acknowledged.



\begin{thebibliography}{10}
\bibitem{covariance}Assuming that the states 1 and 2 of subsystems $A$ and $B$ correspond to the values $a_1$ and $a_2$ of the random variable, we have that the {\it covariance} equals $(a_1-a_2) \kappa$, and the {\it correlation coefficient} equals $\kappa /[p(1-p)]$.
\bibitem{observation}As previously mentioned, it is as a simple illustration that we imposed $S_q(2)=2S_q(1)$ instead of say $S_{2-q}(2)=2S_{2-q}(1)$. The results would then obviously be the same with $(1-q) \leftrightarrow (q-1)$. Consequently, we would have additivity for $1 \le q \le 2$, instead of $0 \le q \le 1$. The  $(1-q) \leftrightarrow (q-1)$ ``duality" appears naturally in nonextensive statistical mechanics (see, for instance, \cite{duality}).
\bibitem{marsh}J. Marsh and S. Earl (work in progress) noticed and kindly communicated to us that, for the present $\kappa$-model, there were also $\kappa >0$ solutions, and also that the additivity of the $ q \ne 1$ entropy $S_q(N)$ was limited to values of $N$ that only achieved infinity for $p=1$.
\bibitem{singular}This is due, as well known, to the fractal structure of the correlation clusters existing at critical points. An instructive description  in nonextensive terms of such special situations has been recently advanced in \cite{robledo}.
\end{thebibliography}
\end{document}